\documentstyle[12pt]{article}
\begin{document}

\title{\large\bf{PHYSICAL MODEL OF THE FLUCTUATING VACUUM \\
AND THE PHOTON AS ITS ELEMENTARY EXCITATION}}
\author{Josiph Mladenov Rangelov, \\
Institute of Solid State Physics , Bulgarian Academy of \\
Sciences , blv.Tzarigradsko Chaussee 72 ,Sofia 1784 , Bulgaria}
\date{}
\maketitle
\begin{abstract}
  A physical model of the fluctuating vacuum (FlcVcm) and the photon as an
elementary collective excitation in a solitary needle cylindrical form are
offered. We assume that the FlcVcm is consistent by neutral dynamides, which
are streamlined in a close-packed crystalline lattice. Every dynamide is a
neutral pair, consistent by massless opposite point-like elementary electric
charges (ElmElcChrgs): electrino (-) and positrino (+). In an equilibrium
position two contrary Pnt-Lk ElmElcChrgs within every one dynamide are very
closely installed one to another and therefore its aggregate polarization and
its ElcFld also have zero values. However the absence of a mass in a rest of
an electrino and positrino makes possible they to display an infinitesimal
inertness of their own QntElcMgnFlds and a big mobility, what permits them to
be found a bigger time in an unequilibrium distorted position. The aggregate
ElcFld of dynamide reminds us that it could be considered as the QntElcFld of
an electric quasi-dipole because both massless electrino and positrino have
the same inertness. The aggregate ElcFld of every dynamide polarizes nearest
neighbour dynamides in an account of which they interact between them-self,
on account of which their photons display a wave character and behaviour. In
order to obtain a clear physical evidence and true physical explanation of an
emission and absorption of RlPhtns, I use Fermi method for the determination
of the time dependence of expansion coefficients of wave function of SchEl in
a hybrid state, using the solution of the Schrodinger quadratic differential
wave equation in partial derivatives with the potentials of Coulomb and of
Lorentz friction force.
\end{abstract}

  A physical model (PhsMdl) of the existent fluctuating vacuum (FlcVcm) and
its elementary excitation photon as a solitary needle cylindrical harmonic
oscillation is offered. It is common known that the physical model (PhsMdl)
presents at us as an actual ingradient of every good physical theory
(PhsThr). It would be used as for an obvious visual teaching the unknown
occurred physical processes within the investigated phenomena. We assume
that the FlcVcm is consistent by neutral dynamides, streamlined in some
close-packed crystalline lattice. Every dynamide is a massless neutral pair,
consistent by two massless opposite point-like (PntLk) elementary electric
charges (ElmElcChrgs): electrino (-) and positrino (+). In a frozen
equilibrium position both opposite PntLk ElmElcChrgs within every dynamide
are very closely installed one to another and therefore the aggregate
polarization of every one dynamide has zero value and its electric field
(ElcFld) also has zero electric intensity (ElcInt). However the absence of a
mass in a rest of the electrino and positrino makes them possible to have a
big mobility and infinitesimal dynamical inertness of its own QntElcMgnFld,
what permits them to be found a bigger time in an unequilibrium distorted
position. The aggregate ElcFld of the dynamide reminds us that it could be
considered as the QntElcFld of an electric quasi-dipole moment
(ElcQusDplMmn) because both opportunity massless electrino and positrino
have the same inertness. For a certain that is why the FlcVcm dos not
radiate real photon (RlPhtn) by itself, as dynamide electric dipole moment
(ElcDplMmn) has a zero value. The aggregate ElcFld of every dynamide
polarizes nearest neighbour dynamides in an account of which nearest
dynamides interact between itself, and in a result of which its elementary
collective excitations have a wave character and behavior. It is richly
clear that the motions in the opposite direction of both opposite PntLk
ElmElcChrgs of an every dynamide creates an aggregate magnetic field
(MgnFld) of every one and the sum of which makes a magnetic part of the free
QntElcMgnFld.

  Although up to the present nobody of scientists distinctly knows are there
some elementary micro particles (ElmMicrPrts) as a fundamental building
stone of the micro world and what the elementary micro particle (ElmMicrPrt)
means, there exists an essential possibility for physical clear and
scientific obvious consideration of the uncommon quantum behavior and
unusual dynamical relativistic parameters of all relativistic quantized
MicrPrts (QntMicrPrts) by means of our convincing and transparent surveyed
PhsMdl. We suppose that the photon is some elementary excitation of the
FlcVcm in the form of a solitary needle cylindrical harmonic oscillation.
The deviations of both PntLk massless opportunity ElmElcChrgs of an every
dynamide from their equilibrium position in the vacuum close-packed
crystalline lattice creates its own polarization, the sum of which creates
total polarization of the FlcVcm as a ideal dielectric, which causes the
existence of a total resultant QntElcFld. Consequently the total
polarization of all dynamides creates own resultant QntElcFld, which is an
electric part of the free QntElcMgnFld. Really, if the deviation of an every
PntLk ElmElcChrg within every one dynamide from its own equilibrium position
is described by dint of formula of collective oscillations (RlPhtns) of
connected oscillators in a representation of second quantization:
\begin{equation}  \label{a}
u_{j}(r)\,=\,\frac{1}{\sqrt{N}}\sum_{q}\sqrt{\frac{\hbar }{2\Theta \omega }}
\,I_{jq}\,\left\{ \,a_{jq}^{+}\exp {\,i(\,\omega t\,-\,qr\,)}
\,+\,a_{jq}\,\exp {-i(\,\omega t\,-\,qr\,)}\,\right\}
\end{equation}

where $\Theta$ is an inertial mass of the electrino and positrino and $%
I_{jq} $ are vector components of the deviation (polarization). If we
multiply the deviation u of every PntLk ElmElcChrg in every dynamide by the
twofold ElmElcChrg value e and dynamide density $W=\frac{1}{\Omega _{o}}$,
then we could obtain in a result the total polarization value of the FlcVcm
within a representation of the second quantization :
\begin{equation}  \label{b}
P_{j}(r)\,=\,\frac{2e}{\Omega_o \sqrt{N}} \sum_{q} \sqrt{\frac{\hbar}{%
2\Theta \omega}}\,I_{jq}\,\left\{\,a_{jq}^{+}\,\exp{\,i\,(\omega t\,-\,qr)}%
\,+\, a_{jq}\,\exp {-\,i\,(\omega t\,-qr\,)}\,\right\}
\end{equation}

  Further we must note that the change of the spring with an elasticity $\chi$
between the MicrPrt and its equilibrium position, oscillating with a
circular frequency $\omega $ by two springs with an elasticity $\tilde{\chi}$
between two MicrPrts, having opportunity ElmElcChrgs and oscillating with a
circular frequency $\tilde{\omega}$ within one dynamide, is accompanied by a
relation $\,2\,\tilde{\chi}\,\simeq\,\chi$. Indeed, if the ,,masses'' of the
oscillating as unharmed dynamide is twice the ,,mass'' of the electrino or
positrino, but the elasticity of the spring between every two neighbor
dynamides in crystaline lattice is fourfold more the elasticity of the
spring between two the MicrPrts, having opportunity ElmElcChrgs and
oscillating one relatively other within one dynamide, while the common
,,mass'' of two the MicrPrts, having opportunity ElmElcChrgs and oscillating
one relatively other within one dynamide is half of the ,,mass'' of the
electrino or positrino. Therefore the circular frequency $\omega $ of the
collective oscillations have well known relation with the Qoulomb potential
of the electric interaction (ElcInt) between two opportunity massless PntLk
ElmElcChrgs electrino and positrino and their dynamical inertial ,,masses''
which can be described by dint of the equations :
\begin{equation}  \label{c1}
\tilde{\omega}^{2}=2\frac{\tilde{\chi}}{\Theta}\quad{\rm and}\quad
\omega^{2}\,= \,\frac{4\chi }{2\Theta }\quad {\rm consequently}\quad \omega
^{2}\,= \,2 \tilde{\omega}^{2}
\end{equation}

\begin{equation}  \label{c2}
\quad {\rm and}\quad {\rm therefore}\quad \Theta \omega ^{2}=\frac{4e^{2}} {%
4\pi \Omega _{o}\,\varepsilon _{o}}\quad {\rm or}\quad \Theta C^{2}\,= \,%
\frac{e^{2}}{4\pi \Omega _{o}\,q^{2}\,\varepsilon _{o}}
\end{equation}

where
\begin{equation}  \label{d}
\,N\Omega _o\,=\,\Omega \quad {\rm and} \quad d = W\,e\,E \quad {\rm or}
\quad \,E = \frac{d}{\Omega _o\varepsilon _o} = \frac{P}{\varepsilon _o}
\end{equation}

we could obtain an expression for the ElcInt of the QntElcMgnFld, well known
from classical electrodynamics (ClsElcDnm) in a representation of the second
quantization:
\begin{equation}  \label{e}
E_{j}(r)=\sum_{q}\sqrt{\frac{2\pi\hbar\omega}{\Omega\varepsilon_{o}}}\,
I_{jq}\,\left\{ \,a_{jq}^{+}\,\exp {i(\omega t-\,qr)}\; +\;a_{jq}\,\exp {%
-i(\omega t-\,qr)}\right\}
\end{equation}

  By dint of a common known defining equality :
\begin{equation}  \label{f}
E_j = -\,\frac{\partial A_j}{\partial t}
\end{equation}

  From (\ref{f}) we could obtain the expression for the vector-potential A of
the QntElcMgnFld in the vacuum in a representation of the second
quantization :
\begin{equation}  \label{g1}
A_{j}(r)=i\,\sum_{q}\,\sqrt{\frac{2\pi\hbar}{\Omega\omega\varepsilon_{o}}}\,
I_{jq}\,\left\{ \,a_{jq}^{+}\,\exp {\,i(\,\omega t-qr\,)}-\,a_{jq}\, \exp{%
-\,i(\,\omega t-qr)}\right\} \,
\end{equation}

or
\begin{equation}  \label{g2}
A_{j}(r)=i\,\sum_{q}\,\sqrt{\frac{2\pi\hbar\omega\mu_{o}}{\Omega q^{2}}}\,
I_{jq}\,\left\{ \,a_{jq}^{+}\,\exp {\,i\,(\omega t-qr\,)}\,-\,a_{jq}\, \exp{%
-\,i\,(\omega t-qr)}\right\}
\end{equation}

  Further by dint of the defining equality $\mu _{o}H=rotA$ from (\ref{g1})
and (\ref{g2}) we could obtain an expression for the MgnInt of QntElcMgnFld,
well known from ClsElcDnm in a representation of the second quantization:
\begin{equation}  \label{h}
H_{j}(r)=\sum_{q}\,\sqrt{\frac{2\pi \hbar \omega }{\Omega \mu _{o}}}\, \left[%
\,\vec{n}_{q}\times I_{lq}\,\right] _{j}\left\{ \,a_{jq}^{+}\, \exp{%
\,i(\,\omega t-qr)}+a_{jq}\,\exp {-\,i(\,\omega t-qr)}\right\}
\end{equation}

where $\vec n_k$ is unit vector, determining the motion direction of the
free QntElcMgnFld. By means of presentation (\ref{h}) of MgnInt $H_j(r)$ and
taking into consideration that $\vec n_q$ is always perpendicular to the
vector of the polarization $I_{jq}$ we obtain that :
\begin{equation}  \label{ha}
\,\left[\,\vec v\,\times \left[\,\vec n_q \times \,I_{jq}\right]\,\right] =
\,\vec n_q\,(\vec v\,\cdot\,I_{jq}\,) - I_{jq}\,(\,\vec v\,\cdot\,\vec n_q\,)
\end{equation}

  From this equation (\ref{ha}) we could understand that if the velocity $v$
of the interacting ElcChrg is parallel of the direction $\vec{n}_{q}$ of the
motion of the free QntElcMgnFld, then the first term in the equation (\ref
{ha}) will been nullified and the second term in the equation (\ref{ha})
will determine the force, which will act upon this interacting ElcChrg. But
when the velocity $\vec{v}$ of the interacting ElcChrg is parallel of the
direction $I_{jq}$ of the motion in the opposite directions of two PntLk
ElmElcChrg of the electrino and positrino and one is a perpendicular to the
direction $\vec{n_{q}}$ of the motion of a free QntElcMgnFld, then the
second term in the equation (\ref{ha}) will be nullified and the first term
in the equation (\ref{ha}) will describe the force, which acts upon this
interacting PntLk ElmElcChrg. It turns out that the interaction between
currents of the electrino and positrino, which is parallel to the vector of
a polarization $I_{jq}$ as $(\vec{v}_{j}=\frac{\omega }{\pi }\,\vec{I}_{jq})$%
, with the QntMgnFld of the free QntElcMgnFld determines the motion and its
velocity of same this free QntElcMgnFld. Indeed, it is well known that the
change of a magnetic flow $\Phi $ creates a ElcFld. Therefore by dint of a
relation (\ref{ha}) we can obtain the following relation :
\begin{eqnarray}  \label{hb}
F_{j} = \frac{e}{C}\left[ \vec{v}\times \vec{H}\right] =\frac{e}{m\,C\,\omega%
} \left[ \vec{E}\times \vec{H}\right] = \sum_{q}\frac{e^{2}}{m\,C}\, \sqrt{%
\frac{2\,\pi\,\hbar}{\Omega\,\varepsilon_{o}}}\sqrt{\frac{2\,\pi\,\hbar} {%
\Omega\,\mu _{o}}}\epsilon _{jkl}\vec{n}_{j}(\vec{I}_{kq}\cdot\vec{I}_{lq})
\nonumber \\
\left\{\,a_{kq}^{+}\,\exp{i(\omega t - qr)} + a_{kq}\,\exp {-i(\omega t - qr)%
} \,\right\}\left\{\,a_{kq}^{+}\exp{i(\omega t - qr)} - a_{kq}\, \exp{%
-i(\omega t - qr)}\,\right\}
\end{eqnarray}

  Therefore by dint of (\ref{hb}) and defining equations (\ref{e}) and (\ref
{g2}) we can obtain:
\begin{equation}  \label{i1}
\frac{1}{\sqrt{\varepsilon \varepsilon _o}} = v \sqrt{\mu \mu _o}\quad{\rm or%
} \quad {\rm at} \quad \frac{1}{\sqrt{\varepsilon _o}} = C \sqrt{\mu _o}%
\quad {\rm we \quad have} \quad C = v . \sqrt{\varepsilon \mu}.
\end{equation}

It is naturally that when some RlPhtn is moving within the space of some
substance, then supplementary polarization of atoms and molecules appears ,
which delay its moving and slow down its velocity. Indeed in this case the
dielectric constant $\varepsilon$ has a following form:
\begin{equation}  \label{i2}
\varepsilon = 1 + \sum_q \frac{4 \pi n(q)[\omega(q) - \omega_c]}{\left[\,m\,
\{4(\omega(q) - \omega_c)^2 + \tau^2\,\omega(q)^4\}\right]}
\end{equation}

  By means of the upper scientific investigation we understand that the
creation of the QntMgnFld by moving oposete PntLk ElmElcChrgs of electrinos
and positrinos within all dynamides together with their agregate QntElcFld
as two components of one free QntElcMgnFld one secures their motion.
Therefore we should write the momentum of the free QntElcMgnFld by means of
the equation of Pointing/Umov, using the definition equations (\ref{e}) and (%
\ref{h}) :
\begin{eqnarray}  \label{j1}
&P = \frac{\left[ E\times H\right]}{4\pi C^{2}} = \sum_{q}\vec{n}_{q}
\frac{\displaystyle{\hbar \omega}}{\displaystyle{2\Omega \sqrt{\varepsilon_{o}
\mu_{o}}}}(I_{jq}\cdot I_{jq})&  \nonumber \\
&\left\{a_{jq}^{+}\exp{i(\omega t - qr)}\,+\,a_{jq}\,\exp{-i(\omega t - qr)}
\,\right\}\times &  \nonumber \\
&\left\{\,a_{jq}^{+}\exp{i(\omega t - qr)}\,+\,a_{jq}\, \exp{-i(\omega t -
qr)}\right\} &
\end{eqnarray}

or
\begin{equation}  \label{j2}
P = \sum_q \vec{n}_q \frac{\hbar\,\omega}{2\,\Omega\,C} \left\{\,a_{jq}^{+}
a_{jq} + a_{jq} a_{jq}^{+} + a_{jq}^{+} a_{jq}^{+} \exp{2i(\omega t - qr)} +
a_{jq} a_{jq} \exp{-2i(\omega t - qr)}\,\right\}
\end{equation}

or
\begin{equation}  \label{j3}
\bar P = \sum_q \vec{n}_q \frac{\hbar \omega }{\Omega\,C} (n_q + 1/2)
\end{equation}

  It is well known that the ElmMicrPrts behavior would be studied by means of
an investigation of their behaviors after their interaction by already well
known ElmMicrPrts. Therefore we shall describe the properties and behavior
of the real photon (RlPhtn) by means of a new physical interpretation of
results of its emission and absorption from atoms at their excitated
Schrodinger electrons (SchEls) from higher energetic state into lower
energetic state or vice versa transition. In such a way we could understand
the origin of some their name by dint of the physical understanding these
determining processes.

  In a first we begin by supposing that the RlPhtn has a form of a solitary
needle cylindrical harmonic soliton with a cross section $\sigma_1$,
determined by the following equation :
\begin{equation}  \label{k1}
\sigma_1 = \pi \{(\delta x)^2 + (\delta y)^2\} = \pi \left[\frac{C}{\omega} %
\right]^2 = \frac{2}{\pi}\left(\frac{\lambda}{2}\right)^2 ,
\end{equation}

which is determined by Heisenberg uncertainty relations:
\begin{equation}  \label{k2}
(\delta p_x)^2 (\delta x)^2 \simeq \frac{\hbar^2}{4} ; \qquad {\rm and}
\qquad (\delta p_y)^2 (\delta y)^2 \simeq \frac{\hbar^2}{4} ;
\end{equation}

where the dispersions are:
\begin{equation}  \label{k3}
(\delta x)^2 \simeq (1/2)\{\frac{C}{\omega}\}^2 = \{\frac{\lambda}{2\pi}%
\}^2; \qquad(\delta y)^2 \simeq (1/2)\{\frac{C}{\omega}\}^2 = \{\frac{\lambda%
}{2\pi}\}^2;
\end{equation}

  It is well known that the probability $P_{12}$ for a transition par second
of some SchEl under ElcIntAct of extern ElcMgnFld from an eigenstate 1 into
an eigenstate 2 is determined by the following formula:
\begin{equation}  \label{k4}
P_{12} = \frac{4}{3} \cdot \frac{e^2}{\hbar C^3}\; ( \omega_{12} )^3\;
|\;\langle\;1\;|\;r\;|\;2\;\rangle\;|^2
\end{equation}

  As the intensity of the ElcMgn emission I, emitted par second is equal of
the product of the probability $P_{12}$ for a transition par second by the
energy $\hbar \omega_{12} $ of the emitted RlPhtn, then for certain
\begin{equation}  \label{k5}
I = P_{12} \hbar \omega_{12} = \frac{4}{3} \frac{e^2}{C^3} {\omega_{12}}^4
\; |\;\langle \;1 \;|\; r \;|\; 2 \;\rangle \;|^2
\end{equation}

  Really the matrix element $\langle 1|r|2 \rangle$ of the SchEl position is
determined by the product of the probability for the spontaneous transition
of a SchEl from an higher energetic level into a lower energetic level and
the number $(n+1)$ for the emission or the number $n$ for the absorption of
a RlPhtn, where n is the number of the RlPhtns within the external
QntElcMgnFld, which polarizes atom. In our view here we need to note obvious
supposition that the spreading quantum trajectory of the SchEl is a result
of the participating of its well spread (WllSpr) ElmElcChrg in isotropic
three dimensional (IstThrDmn) nonrelativistic quantized (NrlQnt) Furthian
stochastic (FrthStch) circular harmonic oscillations motion (CrcHrmOscsMtn),
which is a forced result of the electric interaction (ElcIntAct) of the
SchEl's WllSpr ElmElcChrg by the electric intensity (ElcInt) of the
resultant resonance QntElcMgnFld of all stochastic virtual photons
(StchVrtPhtns), existing in this moment of time within the area, where it is
moving. In order to understand this uncommon stochastic motion we must
remember the IstThrDmn nonrelativistic classical (NrlCls) Brownian
stochastic (BrnStch) trembling harmonic oscillation motion (TrmHrmOscMtn) (
\cite{JMRa}), (\cite{JMRb}), (\cite{JMRc}), (\cite{JMRd}), (\cite{JMRe}), (
\cite{JMRf}).

  Therefore there is no possibility for a classical Lorentz' electron (LrEl)
to be in a hybrid state, as it must go along one smooth classical trajectory
and therefore it has no possibility to tunneling between two different
quantized orbits. But in a natural result of its quantized stochastic motion
it is turn out that the SchEl repeatedly $(\simeq 10^{6}times)$ goes
(tunnels) through the potential barrier between both stationary states by
dint of the ElcIntAct of the SchEl's WllSpr ElmElcChrg by the ElcInt of the
resultant resonance QntElcMgnFld of all StchVrtPhtns existing in this moment
of the time within the barrier area. Really at these periodic tunnelling of
the SchEl it has a possibility to go from one stationary orbit to another
stationary orbit and back again for the time of the emission or the
absorption of a RlPhtn by a purpose to ensure the periodic alteration of the
atomic ElcDplMmn, constituent by SchEl's WllSpr ElmElcChrg and the ion
ElcChrg. It is quite plainly that it needs the optical resonance to be
observed in a case of a coincidence of the proper circular frequency of
these transitions between both energetic state with the radiation frequency $%
(\omega _{c}=\omega _{2}-\omega _{1})$.

  It is necessary here to remember that the light radiation of a solitary
moving WllSpr ElmElcChrg of the SchEl with acceleration between two
stationar states within Qoulomb potential of atomis nuclear charge is caused
by Lorentz' friction. Therefore Fermi (\cite{EF1}) thought that at the
description of the forced flate oscillating FnSpr ElmElcChrg it is necessary
to take into an account the term of Lorentz' friction because of radiation.
Although such consideration, which is developed by Fermi seventy three years
ago permits us to consider as an alternate transition of the SchEl between
both energy levels so and dumping and increasing of the expansion
coefficients of the orbital wave function of both energy levels (OrbWvFncs) $%
\phi_j$, connecting them in the hybrid state, in a time during its
radiation, I think that we must compare the value of different forces
although they have different result and could be considered in different
mathematical ways. In such a way we could understand not only why RlPhtn has
a solitary needle fashion but and why one is radiated in a single form one
after another, We will also to discuss what is a physical cause for
separation of different processes.

  In the first we wish to display our physical understanding all process
within the emission and absorption phenomena. In this purpose Fermi begin
with the determination of the Lorentz' friction force with the MgnIntAct
force. As it is well known from classical electrodynamic (ClsElcDnm) the
value of the Lorentz' friction force is described by a following equality :
\begin{equation}  \label{p1}
{F_j}^{fr} = -e \cdot E_j = (\frac{e}{C})\,\frac{\partial{A_j}^{fr}}{%
\partial t} = \frac{2}{3} \cdot \frac{e^2}{C^3} \stackrel{\ldots}{r}_j
\end{equation}

  After the substitution of $\stackrel{\ldots}{r}_j $ by means of Newton
equality $m{\ddot r}_j = - e E_j $ the value of Lorentz' friction force
takes a following useful form:
\begin{equation}  \label{p2}
{F_j}^{fr} = - i \frac{2}{3} \frac{e^2}{C^3} \frac{\partial E_j}{\partial t}
= \frac{2}{3} \frac{\omega e^2}{C^3} e E_j
\end{equation}

  It is easy to understand that par unit of time the Lorentz' friction force
produces a work, determined by following equation:
\begin{equation}  \label{p3}
W^{fr} = v_j \cdot F^{fr}_j = \frac{2}{3} \frac{e^2}{C^3}({\dot r}_j \cdot {%
\stackrel{\ldots}{r}}_j) = \frac{2e^2}{3C^3} \frac{d}{dt} \left({\dot r}_j
\cdot {\ddot r}_j \right) - \frac{2e^2}{3C^3}\,\left({\ddot r}_j \cdot {%
\ddot r}_j \right)
\end{equation}

  As after an averaging over time the first term $\frac{2\,e^{2}}{3\,C^{3}}\,%
\frac{d}{dt}\left( {\ddot{r}}_{j}\cdot {\dot{r}}_{j}\right) $ is canceled
and therefore the work of Lorentz' friction force coincidences with the
averaged emission energy :
\begin{equation}
\bar{E}=\frac{2}{3}\frac{e^{2}}{\omega C^{3}}({\ddot{r}}_{j}\cdot {\ddot{r}}%
_{j})  \label{p4}
\end{equation}

  After substitution of ${\ddot r}_j$ by its value, determined by means of
Newton's equation ${\ddot r}_j = \frac{e}{m} E_j$ we can obtain a following:
\begin{equation}  \label{p5}
W^{fr} = - \frac{2}{3} \left(\frac{e^2}{mC^2}\right)^2 C\, \left(E_j \cdot
E_j \right) = - \frac{8\pi}{3} \left(\frac{e^2}{mC^2}\right)^2 \frac{C}{4\pi}
\left(E_j \cdot E_j \right)
\end{equation}

  The equation (\ref{p5}) shows us that work, produced by Lorentz' friction
force per unit tine is equal of the product of Thompson total cross section $%
\sigma =\frac{8\pi}{3}\left(\frac{e^{2}}{mC^{2}}\right)^{2}$ of the
dispersed light (RlPhtns) by the FnSpr ElmElcChrg of some free DrEl and its
Pointing/ Umov's vector $S=\frac{C}{4\pi }\left( E_{j}\cdot E_{j}\right)$.
In my point of view I need to make note that if you wish to understand what
is a physical cause for obtaining so big Thompson total cross section of
disspersion of RlPhts from FnSpr ElmElcChrg of DrEl, instead from its PntLk
ElmElcChrg, then you need to read about the participation of its PntLk
ElmElcChrg in own inner self-consistent Zitterbewegung (\cite{JMRc}), (\cite
{JMRd}).

  The same result can be obtained from (\ref{k5}) if we substitute kinetical
energy of forced osccillating SchEl $m\,(\omega_{12})^2\,|\;\langle\;1\;
|\;r\;|\;2\;\rangle\;|^{2}$ by potential energy of its ElcIntAct $\epsilon
\epsilon_o E^{2}$, which follows from Newton relation and the energy
equality. From this physical clear and mathematical correct investigation we
could understand that the friction of such a free ScrEl is determined by
dispersion of the light, which its FnSpr ElmElcChrg can diffract.

  In the second we can obtain that the force of the MgnIntAct between the
electric current $j = ev$ of the SchEl's ElmElcChrg and the MgnInt of the
free QntElcMgnFld, emitted or absorbed by it, is the $\frac{\hbar\omega}{mC^2%
}$ times smaller than the force of the ElcIntAct between the SchEl's
ElmElcChrg and the ElcInt of same free QntElcMgnFld, emitted or absorbed by
it. Although we begin with the calculation of the value of the work,
produced by the MgnIntAct force for unit of time, determined by following
equation:
\begin{equation}  \label{p6}
W^m = \frac{e}{C} \{\vec v \cdot [ \vec v \times H ] \} = \frac{e}{C} \{ [
\vec v \times \vec v ] \cdot H \} = 0
\end{equation}

  From the equation (\ref{p6}) we can see that the Lorentz' MgnIntAct force
cannot participates in the emission and absorption of RlPhtns. Although many
physicists think so writing : $H_{in} = ( j \cdot A ) = - e ( v_j A_j )$,
than this term describe magnetic interaction, but in reality it is wrong.
Really it is easy to understand that :
\begin{equation}  \label{p7}
H_{in} = - e ( v_j A_j ) = - e \frac{d }{dt} ( r_j A_j ) + e ( r_j \frac{%
dA_j }{dt} ) = - e ( r_j E_j )
\end{equation}

  Hence in a reality the Hamiltonian $H_{in}$ describes the ElcIntAct between
the ElcChrg of the particle and the ElcInt of the QntElcFld. Therefore in a
purpose to describe the emission and absorption of RlPhtn we can use two
type of interaction in two different role. The first, ElcIntAct as the
cause, creating the alternate ElcDplMmn by means of which the RlPhtn is
created or absorbed. The second, Lorentz' friction force as the cause, which
restricts the value of the forcing influence of the ElcIntAct at creating
the ElcDplMmn. This physical clear way can be written in a following
mathematical form :
\begin{equation}
m{\ddot{r}}_{j}-m\tau \stackrel{\ldots }{r}_{j}+m{\omega _{c}}%
^{2}r_{j}=-eE_{j}  \label{l1}
\end{equation}

where the first term describes the inertial force, the second term describes
the friction force and the third term describes the elastic force, forcing
the emitting or absorbing SchEl to move between two energetic levels $E_{n}$
and $E_{s}$. The fourth term describes the electric force of ElcIntAct of
the ElcInt of the external QntElcMgnFld by WllSpr ElmElcCrgh of the emitting
or absorbing SchEl. In this way I shall take into consideration most of an
acting interactions by dint of compensation of acting forces at a
determination of radius deviation and only the Lorentz' friction force and
ElcInt by Qoulomb potential. Therefore I shall find a solution of the
inhomogeneous equation (\ref{l2}) in order to determine the time dependence
of the radius deviation r as a result of ElcInt of WllSpr ElmElcChrg of the
emitting or absorbing SchEl with ElcInt of the existent StchVrtPhtn. In this
connection I shall also write into quadratic differential wave equation in
partial deviations of Schrodinger only the Fermi's potential of the Lorentz'
friction force and the Qoulomb potential in same method as the method of
Fermi.

  With a purpose for easily understanding the way of the mathematical solution
of equation (\ref{l1}) we must note that Fermi has ignored three forces: the
inertial force, the elastic force and the electric force, as he thought that
only the Lorentz' friction force transforms emitting SchEl's kinetic energy
to energy of emitted RlPht. Therefore only the Lorentz' friction force
affects over time dependence of expansion coefficients of total wave
function of the emitting SchEl's into a hybrid state during the time of
emission. Therefore for a curtain he has not written and used the solution
of a following inhomogeneous equation :
\begin{equation}  \label{l2}
{\ddot{r}}_{j}=-\frac{e}{m}\,E_{j}
\end{equation}

  For this purpose Fermi have used the Fermi potential $V^{fr}=-\frac{2}{3}%
\frac{e^{2}}{C^{3}}(r_{j}\stackrel{\ldots }{r}_{j})$ of the Lorentz'
friction force of the free moving SchEl instead the electric field potential
$V^{ef}=e(r_{j}E_{j})$ of the oscillating moving SchEl within the
QntElcMgnFld of RlPhtn or VrtPhtn. For a description of the emission and
absorption of photons (RlPhtn and VrtPhtn) we can use the OrbWvFnc $\phi
_{n}(r)$, which are eigenfunction of the time dependent quadratic
differential wave equation within partial derivatives (QdrDfrWvEqtPrtDrv) of
Schrodinger, having a following form :
\begin{equation}
i\hbar \frac{\partial \Psi (r,t)}{\partial t}=-\frac{\hbar ^{2}\Delta }{2m}%
\Psi (r,t)+V_{c}(r)\Psi (r,t)+V_{fr}(r)\Psi (r,t)+V_{ef}(r)\Psi (r,t)
\label{m1}
\end{equation}

where $V_{c}$ is a Coulomb potential, $V_{fr}$ is Lorentz friction potential
and $V_{ef}$ is a potential of the SchEl within the QntElcFld. I think that
it is very interesting that Fermi didn't use known expression for emission
of the electric dipole moment of the emitting atom. Therefore in a real case
the energy of interaction may be presented as potential in a following form:
\begin{equation}
V_{ef}=-\frac{e}{C}(v_{j}.A_{j})=-e(r_{j}.E_{j})  \label{p8}
\end{equation}

  We are use the Newton first motion equation (\ref{l2}) $m {\ddot r}_j = - e
E_j$ in order to determine $r_j$ as a function of $E_j$ by its equal product
$- \frac{e} {m \omega^2} E_j $. In this fashion the Fermi'potential of the
Lorentz' friction force accepts a following form :
\begin{equation}  \label{m2}
V_1 (r) = - \frac{2 e^2}{3 mC^3} (r_j {\dot E}_j)
\end{equation}

  But in a reality we shall use Fermi potential of the Lorentz' friction force
only for its restrictive influence. But time dependence of emitting SchEl's
radius vector $r_{j}$ for description of Lorentz friction force at resonance
case we shall describe by the equation (\ref{l2}) at determination of
emitting SchEl's radius vector $r_{j}$ as a function of the ElcInt of the
external QntElcMgnFld. I must remember you, that $E_{j}$ is the electric
intensity (ElcInt) of the external QntElcFld, within which is found
polarized atom, composed by ion and the charge of the stimulated SchEl. As I
carry out over the potential $V_{1}$ same operation integrating by part,
then we can obtain its other presentation :
\begin{equation}  \label{m3}
V_{1}(r)=-\frac{2}{3}\,\frac{e^{2}}{mc^{3}}\,\frac{d}{dt}(er_{j}E_{j})\,+\,
\frac{2}{3}\,\frac{e^{2}}{mC^{3}}(e{\dot{r}}_{j}E_{j})\,=\, \frac{2}{3}\frac{%
e^{2}}{mC^{3}}(J.E)
\end{equation}

  The expression (\ref{m3}) show us that the potential $V_{1}(r)$ is equal of
the product of the factor of the electric power $W_{1}=(JE)$, which loses a
free SchEl, creating electric current $J$ under the QntElcFld ElcIntAct,
multiplied by the emission time $\tau =\frac{2}{3}\frac{e^{2}}{mC^{3}}$ of
one real photon (RlPhtn). May be therefore Fermi didn'tused the potential $%
V_{1}$ for describing the spontaneous emission of a RlPhtn. Indeed, Fermi
had used the first potential form, from which we could directly obtain
Lorentz friction force ${F^{fr}}_{j}$ :
\begin{equation}
V_{fr}(r)=-\frac{2}{3}\frac{e^{2}}{C^{3}}(r_{j}\stackrel{\ldots }{r}_{j})
\label{m4}
\end{equation}

  In the beginning we determine the orbital wave eigenfunction (OrbEgnWvFnc)
$\Psi_n(r)$ of SchEl within Qoulomb potential of the nuclear electric
charge (NclElcChrg). In its first work (1927) although both potentials
$V_{c}(r)$ and $V_{fr}(r)$ have no time dependence Fermi had found the
solution of the time depending QdrDfrWvEqtPrtDrv of Schrodinger, having a
following form :
\begin{equation}
\label{m5} - \frac{\hbar^2 \Delta} {2.m} \phi_n (r,t) +
V_c (r) \phi_n (r,t) = E_n \phi_n (r,t) ;
\end{equation}

  In order to obtain the total OrbWvFnc $\Psi_{ns} (r,t)$, which is an
eigenfunction of emitting SchEl, moving within Qoulomb potential of the
NclElcChrg:
\begin{equation}  \label{m6}
- \frac{\hbar^2 \Delta}{2.m} \Psi_{ns} (r,t) + V_c (r) \Psi_{ns} (r,t) +
V_{fr} (r) \Psi_{ns} (r,t) = E_n \Psi_{ns} (r,t) ;
\end{equation}

we shall expand the total OrbWvFnc $\Psi _{ns}(r,t)$ in series of the eigen
OrbWvFncs $\phi _{n}(r,t)$ :
\begin{equation}
\Psi _{ns}(r,t)=\sum_{s}\lambda _{s}\phi _{s}\exp {-i\omega _{s}t}
\label{m7}
\end{equation}

  It is very interesting that Fermi had well known, that the electric dipole
moment (ElcDplMmn) $d_{j}=-er_{j})$ of the atom must be time dependent, as
we could see from expression (\ref{m7}). Therefore although he didn't took
into consideration the existence of the time dependent ElcInt $E_{j}$ of the
external QntElcFld, one had artificially put in an use a time dependence of
radius-vector matrix elements. In such an artificial way Fermi written
potential of the Lorentz friction potential (\ref{m3}) in a following
fashion :
\begin{equation}  \label{m8}
\langle V_{fr}\rangle =i\frac{2e^{2}}{3C^{3}} \sum_{n,s} \lambda_{n}^{\ast}
\lambda_{s}(\langle r_{j}\rangle \cdot \langle r_{j}\rangle )\cdot ({\omega}%
_{n} - {\omega }_{s})^{3} \exp {i(\omega _{n}-\omega _{s})t}
\end{equation}

  For calculation of the radius-vector matrix elements $\langle r_j \rangle$
Fermi had used OrbWvFunc $\Psi (r)$, expanded in a power of eigen
OrbWvEgnFncs ${\phi}_n$ of a SchEl, moving within Qoulomb potential of the
NclElcChrg $V_q$ with eigen energy $E_n = \hbar \omega_n$:
\begin{equation}  \label{n1}
\Psi (r,t) = \sum_n \lambda_n \phi_n | n_n \rangle \exp{-i\omega_n t}
\end{equation}

  Therefore Fermi had used OrbWvFunc $\Psi (r)$ of the calculation of the
radius-vector matrix elements $\langle r_{j}\rangle $ in a following fashion
:
\begin{equation}  \label{n2}
\langle r_{j} \rangle = \int \Psi ^{\ast }(r,t)r_{j}\Psi (r,t)d^{3}r =
\sum_{n,s} {\lambda ^{\ast}}_{n} \lambda _{s}{r}_{n,s} \exp {i(\omega_{n} -
\omega _{s})t}
\end{equation}

  Taking into consideration the expression (\ref{m7}), we can put the
expansion (\ref{n1}) into QdrDfrWvEqtPrtDrv of Schrodinger (\ref{m3}) and
obtain the equation for expansion coefficients $\lambda _{n}$ :
\begin{equation}  \label{n3}
\sum_{n}{\dot{\lambda}}_{n}\phi _{n}(r)\exp {i\omega _{n}t} = \frac{i}{\hbar}
\sum_{k}{\lambda }_{n}V_{fr}(r)\phi _{k}(r)\exp {-i\omega _{k}t}
\end{equation}

  In order to obtain the algebraic number equation, depending only from time
in first we must multiply the both sides of the equation (\ref{m3}) by the
SchEl's complex conjugate OrbWvEgnFnc $\phi^{\ast}_n(r)$, multiplied by $%
d^{3}r$, and in second to integrate over the whole space around. Taking into
consideration the orthonormality of the system from the orthogonal and
unitary OrbWvEgnFnc $\phi(r)_{n}^{\ast}$, we can obtain a following system
of equations for the expansion coefficients $\lambda_{n}$ :
\begin{equation}  \label{n4}
{\dot{\lambda}}_{n}= \frac{i}{\hbar } \sum_{s} \lambda_{s} \int {\phi^{\ast}}%
_{n}(r) {V}_{fr}(r) \phi_{s}(r)d^{3}r \exp {i(\omega_{s} - \omega_{n})t}
\end{equation}

  Substituting the matrix element ${\langle V_{1}\rangle }$ by its expression
(\ref{n2}) into equations (\ref{n4}) we could obtain more simple equations :
\begin{equation}  \label{n5}
{\dot{\lambda}}_{n}=-\frac{2}{3}\frac{e^{2}}{\hbar C^{3}} \sum_{l,k,s}
\lambda_{l} \lambda_{k}^{\ast} \lambda_{s}\cdot (\omega_{l} -
\omega_{k})^{3} \cdot \{{\langle l | r_{j} | k \rangle } \cdot {\langle s |
r_{j} | n \rangle}\} \cdot \exp {i(\omega_{k} - \omega_{l} + \omega_{n} -
\omega_{s})t}
\end{equation}

  Secular indignation of the expansion coefficients $\lambda_{k}$ values are
determined by the parts of second power, for which the exponential factors
are came to constants. For obtaining this purpose it is need one to satisfy
a following equality: $\omega _{l}-\omega _{k}+\omega _{s}-\omega _{n}=0$.
If we assume that the rational relations have no between frequencies $%
\omega_{k}$, then this equality is equivalent of other two equalities: $l=n$
and $k=s$. Then taking into consideration only the secular indignation we
could obtain very simply equations :
\begin{equation}  \label{n6}
{\dot{\lambda}}_{n}=-\frac{2}{3}\frac{e^{2}}{\hbar C^{3}} \sum_{s}
\lambda_{n} \lambda_{s}^{\ast }\lambda_{s}(\omega_{n}-\omega_{s})^{3} \cdot
\,\left\{ \,\langle n|r_{j}|s\rangle \,\right\} \cdot \,\left\{ \,\langle s
| r_{j} | n \rangle \,\right\}
\end{equation}

  In further we will investigate a case of emission only of one spectral line,
when all expanding coefficients $\lambda_j$ have zero values with the
exception for an instance of $n=1$ and $s=2$. In this case after accept of a
need used designations :
\begin{equation}  \label{n7}
A = \frac{2}{3}\frac{e^{2}}{\hbar C^3}\,\times\,(\omega _{2}-\omega
_{1})^{3} \cdot \{{\langle 1 |r_{j}| 2 \rangle} \cdot {\langle 2 |r_{j}| 1
\rangle}\} = \frac{4}{3} \frac{e^{2}}{\hbar C} \cdot \frac{\,m\,(\omega_{2}
- \omega_{1})^2\, \{{\langle 1 | r_{j} | 2 \rangle} \cdot {\langle 2 | r_{j}
| 1 \rangle}\}}{2} \cdot \frac{(\omega_{2} - \omega_{1})}{mC^{2}}
\end{equation}

  As it follows from determination (\ref{n7}) the constant A is determined by
the product of the fine structure constant $\alpha = \frac{e^2}{\hbar C}$
with the circular frequency $\omega_c = (\omega_2 - \omega_1)$ and with the
product of the ratio of the quadrate of the circular frequency $(\omega_2 -
\omega_1) = \omega_c$ to the quadrate of the light velocity $C$ (which is an
equal of the quadrate of wave number $q_{c}=(q_{2}-q_{1})$) with the $\frac{2%
}{3}$ of the module quadrate of the matrix element $| \langle n | r_{j} | s
\rangle |^2$ of the radius vector $r_{j}$ of SchEl. In such a way we could
understand that constant $A$ has a inverse of time dimension.

  In second presentation we should see that constant A is a product of the
fine structure constant $\alpha = \frac{e^{2}}{\hbar C},$ the ratio of the
twofold kinetical energy $\frac{4}{3} \frac{e^{2}}{\hbar C} \cdot \frac{%
\,m\, (\omega_{2} - \omega_{1})^{2}\,\{{\langle 1 | r_{j} | 2 \rangle }
\cdot {\langle 2 | r_{j} | 1 \rangle }\}}{2}$ of oscillating SchEl to its
total energy $mC^{2}$ and the circular velocity $(\omega _{2}-\omega _{1})$.
In such a way we could understand that constant $A$ has a inverse of time
dimension.

  In further we can obtain from the equations (\ref{n6}) a following simple
equations by dint of substitution of constant A:
\begin{equation}  \label{o1}
{\dot \lambda}_1 = A \lambda_1 \lambda_2 {\lambda^*}_2 \qquad {{\dot \lambda}%
^*}_1 = A {\lambda^*}_1 \lambda_2 {\lambda^*}_2
\end{equation}

and
\begin{equation}  \label{o2}
{\dot \lambda}_2 = - A \lambda_2 \lambda_1 {\lambda^*}_1 \qquad {{\dot
\lambda}^*}_2 = - A {\lambda^*}_2 \lambda_1 {\lambda^*}_1
\end{equation}

  It is easy for us to see that by multiplying each equation by its complex
conjugated factor and after this by summing of all such obtained new
equations we can obtain their first integral:$\lambda _{1}{\lambda^*}_{1}\,+
\,\lambda _{2}{\lambda ^{\ast }}_{2} = 1$; This result is obvious and very
easy for physical understanding. It shows us that the probability for
finding the emitting SchEl in both energetic levels is preserved during the
emission time.

  It is easy for us to see also that by multiplying of each equation by a its
complex conjugated factor, after summing of both such obtained new pair
equations and after using by substitution obtained first integral we can
obtain following two equations :
\begin{equation}  \label{o3}
\frac{d}{dt} |\lambda_1|^2 = 2 A |\lambda_1|^2 ( 1 - |\lambda_1|^2 ) \qquad
\frac{d}{dt} |\lambda_1|^2 = - 2 A |\lambda_2|^2 ( 1 - |\lambda_2|^2 )
\end{equation}

  After integrating of equations (\ref{o3}) we can obtain following two
solutions:
\begin{equation}  \label{o4}
2 A t = \ln \{ |\lambda_1|^2 /( 1 - |\lambda_1|^2 )\} + const_1\qquad{\bf and%
} \qquad 2 A t = \ln \{ |\lambda_2|^2 /( 1 - |\lambda_2|^2 )\} + const_2
\end{equation}

  If we suppose that $\lambda _{2}=1$ and $\lambda _{1}=0$ at $t=-\infty $
and $\lambda _{2}=0$ and $\Lambda _{1}=1$ at $t=+\infty $, then we can
determine both constant values and obtain following definition equations:
\begin{equation}
|\lambda _{1}|^{2}=\exp ({2At)}\{\exp {2At}+1\}^{-1}\qquad {\bf and}\qquad
|\lambda _{2}|^{2}=\exp ({-2At)}\{\exp {-2At}+1\}^{-1}  \label{o5}
\end{equation}

  In order to obtain their product $|{\lambda ^{\ast }}_{1}\lambda _{2}|$ we
must in first multiply their values from equation (\ref{o5}) and in second
take square from this factor. In such an elementary easy way we can obtain a
simple useful results:
\begin{equation}
|{\lambda ^{\ast }}_{1}\lambda _{2}|={2cosh^{-1}(At)}  \label{o6}
\end{equation}

  The obtained result (\ref{o6}) shows that the real photon (RlPhtn) has a
solitary needle package form of a length l = \{C/A\} of cylindrical harmonic
oscillations, who is emitted for a limited from $A^{-1}$ time. This result
gives very obvious fashion of the RlPhtn, which explain the physical cause,
ensuring the existence of Plank's rule for emission and absorption of every
RlPhtns singly in a solitary needle form. It is a very clear way for correct
obtaining of such physically clear result, but in a reality it don't contain
a very important product of the resonance term. Indeed, as we could see
Fermi has used the Lorentz friction potential $V^{fr}=-\frac{2}{3}\frac{e^{2}%
}{C^{3}} (r_{j}\stackrel{\ldots }{r}_{j})$ of the free moving SchEl instead
the electric field potential (\ref{p7}) $V^{ef}=e(r_{j}E_{j})$ of the
oscillating moving SchEl within the QntElcMgnFld of the existent RlPhtn or
VrtPhtn. It is true as only the friction term would turn the kinetic energy
of excited SchEl in energy of RePhtn.

  Although the obtained results by Fermi is physicaly clear and mathematicaly
correctly nobody turn necessity attention and therefore after the
publication of the work (\cite{WHWP}) by Heisenberg and Pauli, Fermi himself
used their method in works (\cite{EF2}). As I think that Fermi method is
physicaly clear and mathematicaly correctly and gives more obvious picturial
description, I shall thry to use its and deternining the radius vector value
of SchEl by means of equation of motion of forced oscillator with Lorentz
friction as in (\cite{WH}) and (\cite{MB}). In order to obtain better and
more mathematically correct and physically clear solution in the first we
shall use the Lorentz' friction force using the determination of the
deviation radius value $r_j$, created by the influence of the ElcIntAct,
using not only inertial force within the left hand-side of Newton first
motion equation (\ref{l2}) and after that we shall use another presentation
of same Lorentz frictional potential. Really it will be better to use the
Newton first motion equation (\ref{l1}) instead simplified equation (\ref{l2}%
) for determination the radius vector $r_j$ value. In the first we shall
take into account in first the term describing the inertial force, and in
second the term describing the friction force and in third the term
describing the elastic force, which forces the emitting or absorbing SchEl
to move between two energetic levels $E_n$ and $E_s$. In second, we shall
take into consideration the fourth term, describing the electric force of
ElcIntAct of the ElcInt of the external QntElcMgnFld by WllSpr ElmElcCrgh of
the emitting or absorbing SchEl, which secures being of the emitting SchEl
in a hybrid state of two energetic levels, ensures the existence of the
deviation value $r_j$, which determines the electric dipole moment value of
the emitting atom. I reaped, with a purpose for easily and obvious
understanding the way of the mathematical solution of the quadratic
differential wave equation of Schrodinger Fermi has ignored the inertial
force, the elastic force and the electric force, as he thought that only
Lorentz' friction force transforms the kinetic energy of emitting SchEl's to
energy of emitted RlPht. In order to obtain our purpose we shall use the
most correctly equation of motion (\ref{l1}). In this way we shall take into
consideration all terms of interactions and of physical cause, ensuring the
existence of the electric dipole moment (ElcDlpMmn), which ensures the
emission and absorption by atom of some real photons. I think that I am need
to note here that Fermi method is founded aggregate of well considered
expressions. Therefore we must take into account in first the electric
frictional potential of SchEl within external QntElcMgnFld and in second all
terms in the motion equation (\ref{l1}), determining the radius vector $r_j$
as a function of the ElcInt $E_j$. Therefore we shall use the presentation
of the friction potential $V_{fr}$ of SchEl moving within external
QntElcMgnFld:
\begin{equation}  \label{m9}
i \hbar \frac{\partial \tilde \Psi (r,t)}{\partial t} = - \frac{{\hbar}^2
\Delta}{2.m} \tilde \Psi (r,t) + V_c(r) \tilde \Psi (r,t) + V_{fr}(r) \tilde
\Psi (r,t) + V_{ef}(r) \tilde \Psi (r,t)
\end{equation}

 I need to point here, that the influence of the potential $V_{ef}(r)$ within
eqt.(\ref{m9}) will be taken into account by dint of the motion eqt.(\ref{l1}%
). Therefore the influence of Lorentz friction force will be taken into
account by me in same way that Fermi took into account the same Lorentz'
friction force in an approach, neglecting the influence of the potential $%
V_{ef}(r)$. In order to take into account the Lorentz friction force, acting
over emitting SchEl, taking into account the potential $V_{ef}(r)$, its
radius vector $r_{j,q}$ must be determined by means of classical Newton
motion equation (\ref{l1}), taking into account the inertial force, the
frictional force, the elastic force and the electric force from eqt.(\ref{p8}%
), instead by means of Newton motion equation (\ref{l2}), if the ElcInt $E_j$
of the external QntElcMgnFld could be determined by dint of the eqt (\ref{e}%
). In such a way we can obtain a following presentation:
\begin{eqnarray}  \label{q1a}
r_{j,q} = (\frac{e}{m})\,\sqrt{\frac{2\,\pi\,\hbar\,\omega}{\Omega\,
\varepsilon_o}}\,I_{j,q}\, \left\{\frac{\displaystyle{\,a^{+}_{j,q}\, \exp{%
i(\omega t - q.r)}\,}}{\displaystyle{\{\omega^2 - (\omega_2 - \omega_1)^2 +
i\tau \omega^3\}}}\, +\, \frac{\displaystyle{\,a_{j,q}\,\exp{-i(\omega t +
q.r)}}}{\displaystyle {\{\omega^2 - (\omega_2 - \omega_1)^2 - i\tau
\omega^3\}}}\, \right\}
\end{eqnarray}

\begin{eqnarray}  \label{q1b}
r^*_{j,q} = (\frac{e}{m})\,\sqrt{\frac{2\,\pi\,\hbar\,\omega}{\Omega\,
\varepsilon_o}}\,I_{j,q}\, \left\{\,\frac{\displaystyle{\,a_{j,q}\, \exp{%
-i(\omega t - q.r)}\,}}{\displaystyle{\{\omega^2 - (\omega_2 - \omega_1)^2 -
i\tau \omega^3\}}}\,+\, \frac{\displaystyle{\,a^{+}_{j,q}\,\exp{i(\omega t -
q.r)} \,}}{\displaystyle {\{\omega^2 - (\omega_2 -\omega_1)^2 + i\,\tau
\omega^3\}}}\, \right\}
\end{eqnarray}

  But these are an operator presentations of forced oscillation radius
$r_{j,q} $ and we must write their matrices presentation in analogous of
(\ref{n2}) :
\begin{eqnarray}
&\langle \tilde{\Psi}^{\ast}|r_{j,q}|\tilde{\Psi}\rangle = \sum_{p,l}\,
\lambda_{p}^{\ast }\,\lambda_{l}\,\exp{\ i(\omega_{p}-\omega_{l})t}(\frac{e}
{m})\,\,\sqrt{\frac{2\,\pi\,\hbar \,\omega }{\Omega \,\varepsilon _{o}}}\;
\times &  \nonumber
\label{q2a} \\ &\left\{\langle \,\phi_{p}^{\ast }|r_{j,o}I_{j,q}\exp
{-i(q.r)}|\phi _{l}\,\rangle \,\times \,\frac{\displaystyle{\langle
n_{p}|a_{j,q}^{+}|n_{l}\rangle \,\exp{i(\omega t)}\,}}{\displaystyle{%
\{\omega^{2}-(\omega_{2}-\omega_{1})^{2}+i\tau \omega^{3}\}}}\,\right. &
\nonumber \\
&+\,\left.\langle \,\phi_{p}^{\ast }|r_{j,o}I_{j,q}\exp{+i(q.r)}|\phi_{l}\,
\rangle \,\times \,\frac{\displaystyle{\langle \,n_{p}|a_{j,q}|n_{l}\,\rangle
\,\exp{-i(\omega t)}\,}}{\displaystyle{\{\omega^{2}-(\omega_{2}-\omega_{1})^{2}
-i\tau \omega ^{3}\}}}\,\right\} &
\end{eqnarray}

\begin{eqnarray}
&\langle \,\tilde{\Psi}^{\ast }|r_{j,q}^{\ast }|\tilde{\Psi}\rangle
=\sum_{p,l}\,\lambda_{p}^{\ast }\,\lambda_{l}\,\exp{i(\omega _{p}-\omega_{l})
t}\,(\frac{e}{m})\,\,\sqrt{\frac{2\,\pi \,\hbar \,\omega }{\Omega
\varepsilon _{o}}}\;\times \,&  \nonumber  \label{q2b} \\
&\left\{ \langle \,\phi _{p}^{\ast }|r_{j,o}I_{j,q}\,\exp {+i(q.r)}|\phi
_{l}\,\rangle \,\times \,\frac{\displaystyle{\langle
\,n_{p}|a_{j,q}|n_{l}\,\rangle \,\exp {-i(\omega t)}\,}}{\displaystyle{%
\{\omega ^{2}-(\omega _{2}-\omega _{1})^{2}-i\tau \omega ^{3}\}}}\,\right. &
\nonumber \\
&+\,\left. \langle \,\phi _{p}^{\ast }|r_{j,o}I_{j,q}\exp {-i(q.r)}|\phi
_{l}\,\rangle \,\times \,\frac{\displaystyle{\langle
\,n_{p}|a_{j,q}^{+}|n_{l}\,\rangle \exp {i(\omega t)}\,}}{\displaystyle{%
\{\omega ^{2}-(\omega _{2}-\omega _{1})^{2}+i\,\tau \omega ^{3}\}}}%
\,\right\} &
\end{eqnarray}

  If $E_{j,q}=\sqrt{\frac{2\pi \hbar \omega }{N\omega _{o}\varepsilon _{o}}}$,
then we could easily verify that $\frac{e}{m}E_{jq}=\sqrt{\frac{4\pi ee}{%
N\omega _{o}\varepsilon _{o}}\frac{\hbar }{2m\omega }}={\omega }^{2}\sqrt{%
\frac{\hbar }{2m\omega }}={\omega }^{2}r_{jo}$, where $r_{jo}$ is an
amplitude of an oscillation of one RlPhtn. Therefore the expression of the
ElcDplMmn matrix element $\langle d_{\omega }\rangle $ has the following
matrix presentation :
\begin{eqnarray}
&\langle \tilde{\Psi}^{\ast }|d_{j,q}|\tilde{\Psi}\rangle =\sum_{p,l}\lambda
_{p}^{\ast }\lambda _{l}\exp {i(\omega _{p}-\omega _{l})t}\,\ \ (\frac{e^{2}%
}{m})\,\sqrt{\frac{2\,\pi \,\hbar \,\omega }{\Omega \,\varepsilon _{o}}}%
\,\times \,&  \nonumber  \label{q3a} \\
&\left\{ \,\langle \,\phi _{p}^{\ast }\,r_{j,o}I_{j,q}\,\exp {-i(q.r)}\,\phi
_{l}\,\rangle \,\times \,\frac{\displaystyle{\langle
\,n_{p}|a_{j,q}^{+}|n_{l}\,\rangle \,\exp {i(\omega t)}\,}}{\displaystyle{%
\{\omega ^{2}-(\omega _{2}-\omega _{1})^{2}+i\tau \omega ^{3}\}}}\;\right. &
\nonumber \\
&+\left. \,\langle \,\phi _{p}^{\ast }\,r_{j,o}I_{j,q}\,\exp {+i(q.r)}\,\phi
_{l}\,\rangle \,\times \,\frac{\displaystyle{\langle
\,n_{p}|a_{j,q}|n_{l}\,\rangle \,\exp {-i(\omega t)}\,}}{\displaystyle{%
\{\omega ^{2}-(\omega _{2}-\omega _{1})^{2}-i\tau \omega ^{3}\}}}\;\right\} &
\end{eqnarray}

\begin{eqnarray}
&\langle \tilde{\Psi}^{\ast }|d_{j,q}^{\ast }|\tilde{\Psi}\rangle
=\sum_{p,l}\lambda_{p}^{\ast }\lambda_{l}\exp{i(\omega_{p}-\omega_{l})}%
\,\ \ (\frac{e^{2}}{m})\,\sqrt{\frac{2\,\pi \,\hbar \,\omega }{\Omega
\,\varepsilon _{o}}}\,\times \,&  \nonumber  \label{q3b} \\
&\left\{\langle \,\phi_{p}^{\ast }\,r_{j,o}I_{j,q}\,\exp{+i(q.r)}\,\phi_{l}\,
\rangle \,\times \,\frac{\displaystyle{\langle n_{p}|a_{j,q}|n_{l}\rangle \,
\exp{-i(\omega t)}\,}}{\displaystyle{\{\omega^{2}-(\omega_{2}-\omega_{1})^{2}-
i\tau \omega^{3}\}}}\;\right. &
\nonumber \\
&+\left. \,\langle \,\phi _{p}^{\ast }|r_{j,o}I_{j,q}\exp{-i(q.r)}|\phi_{l}\,
\rangle \,\times \,\frac{\displaystyle{\langle\,n_{p}|a_{j,q}^{+}|n_{l}\,
\rangle \,\exp{i(\omega t)}\,}}{\displaystyle{\{\omega^{2}-(\omega_{2}-
\omega_{1})^{2}+i\,\tau \omega^{3}\}}}\;\right\} &
\end{eqnarray}

  In order to obtain the total value of the ElcDplMmn matrix element we need
to multiply its value (\ref{q3a},\ref{q3b}) by $\frac{\Omega_o N}{8{\pi}%
^3\,C^3} {\omega}^2 d\omega$ and its product to integrate from zero to $%
\infty$. It will be more suitable instead $r_o I_j $ to write $\langle 0 |
r_j | 1 \rangle$ or $\langle 1 | r_j | 0 \rangle$. As we can understand the
different value from zero will have only the parts, satisfying the
conditions $\langle n_k | a^{+}_{jq} | n_s \rangle \ne 0 $ and $\langle n_k
| a_{jq} | n_s \rangle \ne 0 $, depending from the correlation between $n_k$
and $n_s$. In order to obtain the expansions (\ref{q3a}, \ref{q3b}) we shall
use only two OrbWvEgnFncs $\phi_{1}$ and $\phi_{2}$ if : $H_{o}\phi_{1} =
E_{1}\phi_{1}$ and $H_{o}\phi_{2} = E_{2}\phi_{2}$. For description of the
emission and absorption of photons (RlPhtn and VrtPhtn) we once again can
use the OrbEgnWvFnc $\Psi^{o}(r)$, which are OrbWvEgnFnc of the time
dependent quadratic differential wave equation within partial derivatives
(QdrDfrWvEqtPrtDrv) of Schrodinger, having the form (\ref{m1}). We suppose
that the OrbEgnWvFncs $\phi_{n}^{o}(r)$ of the SchEl within Qoulomb
potential of the nuclear electric charge (NclElcChrg) are determined in the
beginning. Then in a result of using the presentations of $r_{j,q}^{*}$ and $%
E_{j,q}$ within potential (\ref{p7}) we can obtain a following
representation of its matrix element :
\begin{eqnarray}  \label{q2}
&\langle \phi ^{\ast }|V_{fl}|\phi \rangle =\frac{e^{2}}{m}\sum_{q,m,s,n,l}\,
 \lambda_{n}^{*}\lambda_{s}\lambda_{p}^{*}\lambda_{l}(\frac{2\pi \hbar \omega}
{\Omega })\,(\omega_{l} - \omega_{k})^{3}\,\exp{i(\omega_{n} - \omega_{s} +
\omega_{p} - \omega_{l})t}\;\times &  \nonumber \\
&\left[\,\left\{\,\int_{\Omega}\,\phi_{n}^{*}(r)\,\phi_{s}(r)\,I_{j,q}\, \exp%
{+(iq.r)}\,d^{3}\,r\,\,\times\,\frac{\displaystyle{\langle\,n_{n}|
a_{j,q}^{+}|n_{s}\,\rangle\,\exp{(i\omega t)}\,}}{\displaystyle{\
\{\omega^{2} - (\omega _{2} - \omega _{1})^{2} + i\,\tau \omega^{3}\}}}\;
\right\}\,\right. &  \nonumber \\
&\left.\left\{\,\int_{\Omega}\,\phi_{n}^{*}(r)\,\phi_{s}(r)\,I_{j,q}\, \exp{%
+(iq.r)}\,d^{3}\,r\,\times\,\frac{\displaystyle{\langle\,n_{n}| a_{j,q}|
n_{s}\,\rangle\,\exp{-i(\omega t)}\,}}{\displaystyle{\{\omega^{2} -
(\omega_{2} - \omega_{1})^{2} - i\,\tau \omega^{3}\}}}\;\right\} \,\right]\,
\times\,&  \nonumber \\
&\left[\,\left\{\,\int_{\Omega}\,\phi_{p}^{*}(r^{\prime})\,\phi_{l}(r^{%
\prime})\,I_{j,q}\, \exp {+i(q.r^{\prime})}d^{3}r^{\prime}\,\times\,\frac{%
\displaystyle{\langle \,n_{p}| a_{j,q}^{+}|n_{l}\,\rangle\,\exp{i(\omega t)}%
\,}}{\displaystyle{\ \{\omega^{2} - (\omega_{2} - \omega_{1})^{2} + i\,\tau
\omega^{3}\}}}\; \right\}\,\right.&  \nonumber \\
&+\,\left.\left\{\,\int_{\Omega}\,\phi_{p}^{*}(r^{\prime})\,\phi_{l}(r^{%
\prime})\,I_{j,q}\, \exp{+(iq.r^{\prime})}\,d^{3}\,r^{\prime}\,\times\,\frac{%
\displaystyle{\langle\,n_{p}| a_{j,q}| n_{l}\,\rangle\,\exp{-i(\omega t)}\,}%
}{\displaystyle{\{\omega^{2} - (\omega_{2} - \omega_{1})^{2} - i\,\tau
\omega^{3}\}}}\;\right\}\,\right]&
\end{eqnarray}

  In analogous with the eqn (\ref{n3}) we can rewrite eqn(\ref{q2}) in the
equation for new coefficients $\lambda_n$ :
\begin{eqnarray}  \label{q3}
&\sum_s {\dot{\lambda}}_n \int_\Omega\,\phi^*_n(r)\,\phi_s(r)\,d^3\,r\, \exp{%
-i(\omega_s t)} = \frac{2i\pi e^2}{m \Omega}\,\sum_{q,m,s,n,l}\, \lambda_s
\lambda^*_p \lambda_l\,(\omega_{l} - \omega_{k})^{3}\, \exp{i(\omega_p -
\omega_s -\omega_l)t}\,\times\,&  \nonumber \\
&\left[\,\left\{\,\int_\Omega\,\phi^*_n(r)\,\phi_s(r)\,I_{j,q}\,\exp{-(iq.r)}
\,d^3\,r\,\times\,\frac{\displaystyle{\omega\,\langle\,n_n | a^{+}_{j,q} |
n_s\,\rangle\,\exp{i(\omega t)}\,}}{\displaystyle{\{\omega^2 - (\omega_2 -
\omega_1)^2 + i\,\tau \omega^3 \}}}\,\right\}\,\right.&  \nonumber \\
&+\,\left.\left\{\,\int_\Omega\,\phi^*_n(r)\,\phi_s(r)\,I_{j,q}\,\exp{+(iq.r)%
}\, d^3\,r\,\times\,\frac{\displaystyle{\omega\,\langle\,n_n | a_{j,q} |
n_s\, \rangle \,\exp{-i(\omega t)}\,}}{\displaystyle{\{\omega^2 - (\omega_2
- \omega_1)^2 - i\,\tau \omega^3\}}}\,\right\}\,\right]\,\times &  \nonumber
\\
&\left[\,\left\{\int_\Omega\,\phi^*_p(r^{\prime})\,\phi_l(r^{\prime})%
\,I_{j,q}\,\exp{-i(q.r^{\prime})} \,d^3\,r^{\prime}\,\times\,\frac{%
\displaystyle{\omega\,\langle\,n_p | a^{+}_{j,q} | n_l \,\rangle\,\exp{%
i(\omega t)}\,}}{\displaystyle{\{\omega^2 - (\omega_2 - \omega_1)^2 +
i\,\tau \omega^3\}}}\,\right\}\,\right.&  \nonumber \\
&+\left.\left\{\,\int_\Omega\,\phi^*_p(r^{\prime})\,\phi_l(r^{\prime})%
\,I_{j,q}\,\exp{+i(q.r^{\prime})} \,d^3\,r^{\prime}\,\times\,\frac{%
\displaystyle{\omega\,\langle\,n_p | a_{j,q} | n_l\, \rangle\,\exp{-i(\omega
t)}\,}}{\displaystyle{\{\omega^2 - (\omega_2 - \omega_1)^2 - i\,\tau
\omega^3\}}}\,\right\}\,\right]&
\end{eqnarray}

  After taking into consideration the ortogonality of OrbEgnWvFnks $\phi^*_n$
and $\phi_s$ we can translate factor $\exp{-i\omega_n t}$ from the left-hand
side into the right-hand side and after making essential partial
multiplications the expression (\ref{q3}) can been rewritten in a following
form :
\begin{eqnarray}  \label{q4}
&\dot {\lambda}_n = \frac{4i\pi\,e^2}{\hbar\,\Omega}\,\sum_{q,n,s,p,l}\,
\lambda_s \lambda^*_p \lambda_l\;(\omega_{l} - \omega_{k})^{3}\; \exp{%
i(\omega_n + \omega_p - \omega_s - \omega_l)t}\, \;\times\,&  \nonumber \\
&\left\{\int_\Omega\,\phi^*_n(r)\,\phi_s(r)\,r^o_j I_{j,q}\,\exp{-(iq.r)}
\,d^3 \,r
\,\times\,\int_\Omega\,\phi^*_p(r^{\prime})\,\phi_l(r^{\prime})\,r^o_j
I_{j,q}\, \exp{+i(q.r^{\prime})} \,d^3 r^{\prime}\right.&  \nonumber \\
&\times\,\frac{\displaystyle{\omega^2\,(\langle\,n_n | a^{+}_{j,q} | n_s\,
\rangle\,\times\,\langle\,n_p | a_{j,q} | n_l\,\rangle)}}{\displaystyle{%
\{\omega^2 - (\omega_2 - \omega_1)^2 - i\,\tau \omega^3\}}}\;+\;\frac{%
\displaystyle{\omega^2\, (\langle\,n_n | a_{j,q} | n_s\,\rangle\,\times\,
\langle\,n_p | a^{+}_{j,q} | n_l\,\rangle)}}{\displaystyle{\{\omega^2 -
(\omega_2 - \omega_1)^2 + i\,\tau \omega^3\}}}\;\times\,&  \nonumber \\
&\int_\Omega\,\phi^*_n(r)\,\phi_s(r)\,r^o_j I_{j,q}\,\exp{+i(q.r)}\,d^3\,r\,
\times\,\int_\Omega\,\phi^*_p(r^{\prime})\,\phi_l(r^{\prime})\,r^o_j
I_{j,q}\,\exp{-i(q.r^{\prime})}\, d^3\,r^{\prime}&  \nonumber \\
&+\,\int_\Omega\,\phi^*_n(r)\,\phi_s(r)\, r^o_j I_{j,q}\,\exp{(-iq.r)}%
\,d^3\,
r\,\times\,\int_\Omega\,\phi^*_p(r^{\prime})\,\phi_l(r^{\prime})\,r^o_j
I_{j,q} \,\exp{-i(q.r^{\prime})}\,d^3\,r^{\prime}\,\times&  \nonumber \\
&\frac{\displaystyle{\omega^2\,(\langle\,n_n | a^{+}_{j,q} | n_s\,\rangle\,
\times\,\langle\,n_p | a^{+}_{j,q} | n_l\,\rangle)\exp{(+2i\omega t)}}} {%
\displaystyle{\{\omega^2 - (\omega_2 - \omega_1)^2 - i\,\tau \omega^3\}}}%
\;+\; \left.\frac{\displaystyle{\omega^2\,(\langle\,n_n | a_{j,q} |
n_s\,\rangle\, \times\,\langle\,n_p | a_{j,q} | n_l\,\rangle)\exp{-2i(\omega
t)}}} {\displaystyle{\{\omega^2 - (\omega_2 - \omega_1)^2 + i\,\tau
\omega^3\}}}\; \right.&  \nonumber \\
&\left.\times\,\int_\Omega\,\phi^*_n(r)\,\phi_s(r)\,r^o_j I_{j,q}\,\exp{%
+i(q.r)} \,d^3\,r\,\times\,\int_\Omega\,\phi^*_p(r^{\prime})\,\phi_l(r^{%
\prime})\,r^o_j I_{j,q}\,\exp{+i(q.r^{\prime})}\,d^3\,r^{\prime}\right\}&
\end{eqnarray}

  Secular indignation of the expansion coefficients $\lambda_n$ values are
determined by the parts of second power, for which the exponential factors
are came to constants. For obtaining this purpose it is need one to satisfy
a following equality: $\omega_l - \omega_p + \omega_s - \omega_n = 0$. If
there os not the rational relations between frequencies $\omega_n$, then
this equality is equivalent of other two equalities: $l = n$ and $p = s$.
Then taking into consideration only the secular indignation we can obtain
very simply equations :
\begin{eqnarray}  \label{q5}
&\dot {\lambda}_n = \frac{4\,i\,\pi\,e^2}{\hbar\,\Omega}\,\sum_{q,n,s,p,l}\,
\lambda_l \lambda_s \lambda^*_p\;(\omega_{l} - \omega_{n})^{3}\, \exp{%
i(\omega_n + \omega_p - \omega_s - \omega_l)t}\,\times&  \nonumber \\
&\left\{\int_\Omega\,\phi^*_n(r)\,\phi_s(r)\,r^o_j I_{j,q}\,\exp{+i(q.r)}
\,d^3\,r\,\times\,\int_\Omega\,\phi^*_s(r^{\prime})\,\phi_n(r^{\prime})%
\,r^o_jI_{j,q}\, \exp{+i(q.r^{\prime})}\,d^3\,r^{\prime}\right.&  \nonumber
\\
&\times\,\frac{\displaystyle{\omega^2\,(\langle\,n_n | a^{+}_{j,q} | n_s\,\
rangle\,\times\,\langle\,n_s | a_{j,q} | n_n\,\rangle)}}{\displaystyle{\
\{\omega^2 - (\omega_2 - \omega_1)^2 - i \tau \omega^3\}}}\;+\; \frac{%
\displaystyle{\omega^2\, (\langle\,n_n | a_{j,q} | n_s\,\rangle\,\times\,
\langle\,n_s | a^{+}_{j,q} | n_n\, \rangle)}}{\displaystyle{\{\omega^2 -
(\omega_2 - \omega_1)^2 + i\,\tau \omega^3\}}}\;\times\,&  \nonumber \\
&\int_\Omega\,\phi^*_n(r)\,\phi_s(r)\,r^o_j I_{j,q}\,\exp{-i(q.r)}\,d^3\,r\,
\times\,\int_\Omega\,\phi^*_s(r^{\prime})\,\phi_n(r^{\prime})\,r^o_jI_{j,q}%
\,\exp{+i(q.r^{\prime})} \,d^3\,r^{\prime}&  \nonumber \\
&+\,\int_\Omega\,\phi^*_n(r)\,\phi_s(r)\,r^o_j I_{j,q}\,\exp{-i(q.r)}%
\,d^3\,r\,
\times\,\,\int_\Omega\,\phi^*_s(r^{\prime})\,\phi_n(r^{\prime})\,r^o_j
I_{j,q}\,\exp{-i(q.r^{\prime})}\, d^3\,r^{\prime}\times&  \nonumber \\
&\frac{\displaystyle{\omega^2\,(\langle\,n_n | a^{+}_{j,q} | n_s\,\rangle\,
\times\,\langle\,n_s | a^{+}_{j,q} | n_n\,\rangle)\,\exp{+2i(\omega t)}\,}} {%
\displaystyle{\{\omega^2 - (\omega_2 - \omega_1)^2 - i \tau \omega^3\}}}\;+
\;\frac{\displaystyle{\omega^2\,(\langle\,n_n | a_{j,q} | n_s\,\rangle\,
\times\,\langle\,n_s | a_{j,q} | n_n\,\rangle)\,\exp{-2i(\omega t)}\,}} {%
\displaystyle{\{\omega^2 - (\omega_2 - \omega_1)^2 + i\,\tau \omega^3\}}}\; &
\nonumber \\
&\left.\times\,\int_\Omega\,\phi^*_n(r)\,\phi_s(r)\,r^o_j I_{j,q}\,\exp{%
+i(q.r)} \,d^3\,r\,\times\,\int_\Omega\,\phi^*_s(r^{\prime})\,\phi_n(r^{%
\prime})\,r^o_j I_{j,q}\, \exp{+i(q.r^{\prime})}\,d^3\,r^{\prime}\,\right\}&
\end{eqnarray}

  From eqt (\ref{q5}) we see that last two parts are very frequently
alternating in time and therefore after some time averaging they could be
ignored. In these approximation by dint of equations (\ref{q2a}) and (\ref
{q2b}) we can obtain the presentation, known from eqt (\ref{n6}) :
\begin{equation}  \label{q6}
\dot {\lambda}_n = - \frac{2}{3}\,\frac{e^2}{\hbar\,C^3} \sum_s\,\lambda_l\,
\lambda_s^*\,\lambda_s\,\times (\omega_n - \omega_s)^3 \cdot \{ \langle\,n |
r_j | s\,\rangle \}\,\times\,\{ \langle\,s | r_j | n\,\rangle \}
\end{equation}

  The reception of known expressions for the time dependence of coefficients
of the expansion $\lambda_{n}$ of the hybrid state of the emitting or
absorbing SchEl taking into consideration the influence of the ElcInt $E_{j}$
and the oscillating force of a continuous transition between two energetic
level allows us to obtain the resonant form of (\ref{q2a},\ref{q2b}) instead
the simple form (\ref{l2}) of the same radius deviation. After all that we
should understand that the useful and obvious supposition of Fermi about the
physical consequence of the influence of Lorentz friction force is true and
deserves to be used in the explanation of Plank's rule for emission and
absorption of RlPhtn in a solitary needle form. Besides that, there are no
forced or spontaneous emissions and forced absorption. In reality all
emissions are forced but some of them are forced from the electric fields of
real photons and others of them are forced from the electric fields of
virtual photons. In a time of the forced emission and the absorption the
product of both coefficients $\lambda_1$ and $\lambda_2$ of both OrbWvFncs $%
\phi_1$ and $\phi_2$ determines the time dependence of the forced radius
deviation and of the intensity of the emission or the absorption by dint of
the SchEl. Therefore the participation of the SchEl in the process of the
emission or the absorption is determined and limited from the Lorentz'
friction force. Therefore the emitted real photons are quants of the
quantized electromagnetic energy, which have a solitary needle form.

  The reception of known expressions for the ElcInt and MgnInt values of the
QntElcMgnFld by dint of a simple transformation of an expression, describing
deviation of two PntLk ElmElcChrgs of distorted dynamides into the ideal
dielectric of the FlcVcm proves obviously and scientifically the true of our
assumption about the dipole structure of the vacuum and about the creation
way of its collective oscillation - RlPhtn. The existence of a possibility
for a creation of virtual photons (VrtPhtns) as an excitation within the
fluctuating vacuum (FlcVcm) renders an essential influence over the motion
of a electric charged or magnetized micro particles (MicrPrts) by means of
its EntElcMgnFld. The existence of a free energy in the form of micro
particles (MicrPrts) can break of the connection between pair contrary PntLk
ElmElcChrgs of one dynamide and to excite pair of two opposite charged
MicrPrts at once.

  As all MicrPrts are excitements of the vacuum then every one of them would
can move freely through its ideal dielectric lattice without any friction or
damping, that is to say why ones move without to feel the existence of the
vacuum.  Moreover, the existence of some MicrPrt in the easily polarized
FlcVcm distorts its ideal crystalline lattice by influence of its high
dinsity own QntElcMgnFld, created by own FnSpr ElmElcChrg. This natural
distortion of the neutral moleqular FlcVcm with the close-packed lattice
excites and enssures the gravitation field of the ElmMicrPrt's mass, which
by using same force show attention upon mass of another ElmMicrPrt and upon
its behavior. The equivalence of both presentations, of the Coulob and
Newton potentials and forces of interactions is a result of the dimensional
equality of the space, within they act. In such a naturally obvious and
physicaly clear way we understand why the force of the gravitation
interaction is determined by the self energy at a rest and mass.

\end{document}